\documentclass{aipproc}

\layoutstyle{6x9}

\usepackage{graphicx,epsf}
\usepackage{psfrag}

\def\al{&}
\def\be{\begin{equation}}
\def\ee{\end{equation}}
\def\bea{\begin{eqnarray}}
\def\eea{\end{eqnarray}}

\newcommand{\bdm}{\begin{displaymath}}
\newcommand{\edm}{\end{displaymath}}
\newcommand{\no}{\nonumber \\}

\newcommand{\fs}{\; .}
\newcommand{\co}{\; ,}

\newcommand{\QCD}{{\scriptscriptstyle QCD}}
\newcommand{\indV}{{\scriptscriptstyle V}}

\newcommand{\indA}{{\scriptscriptstyle A}}

\newcommand{\indL}{{\scriptscriptstyle L}}
\newcommand{\indR}{{\scriptscriptstyle R}}
\newcommand{\lvac}{\langle 0|\,}
\newcommand{\rvac}{\,|0\rangle}

\newcommand{\qbar}{\overline{\rule[0.42em]{0.4em}{0em}}\hspace{-0.45em}q}
\newcommand{\ubar}{\overline{\rule[0.42em]{0.4em}{0em}}\hspace{-0.5em}u}
\newcommand{\dbar}{\,\overline{\rule[0.65em]{0.4em}{0em}}\hspace{-0.6em}d}

\newcommand{\lbar}{\bar{\ell}}

\newcommand{\ind}{\scriptscriptstyle}

\newcommand{\rs}{\langle r^2\rangle\rule[-0.2em]{0em}{0em}_s}
\newcommand{\lsim}{\,\raisebox{-0.3em}{$\stackrel{\raisebox{-0.1em}{$<$}}{\sim}$}\,}

\begin{document}
\title{Pion Physics at Low Energy and High Accuracy}
      
\author{H. Leutwyler}{email={leutwyler@itp.unibe.ch},
  address={Institute for Theoretical Physics, University of Bern,
  Sidlerstr.~5, CH-3012 Switzerland}}

\begin{abstract}
The role of the quark condensate for the low energy structure of QCD is
discussed in some detail. In particular, the dependence of $M_\pi$
on $m_u$ and $m_d$ and the low energy theorems for the $\pi\pi$
scattering amplitude are reviewed. The new data on 
$K_{e_4}$ decay beautifully confirm the standard picture, according to which
the quark condensate is the leading order parameter of the spontaneously
broken chiral symmetry.
\end{abstract}

\maketitle

\section{1.\hspace{0.6em}Standard Model for 
$\mathbf{E\ll M_{\ind W}}$}
At energies that are small compared to $\{M_W,M_Z,M_H\}=O(100\,\mbox{GeV})$, 
the weak interaction freezes out, because these
energies do not suffice to excite the relevant degrees of freedom.  
As a consequence, the gauge group of the Standard Model, 
SU(3)$\times$SU(2)$\times$U(1), breaks down to
the subgroup SU(3)$\times$U(1) -- only the photons, the gluons, the quarks and
the charged leptons are active at low energies. Since the neutrini
neither carry colour nor charge, they decouple. 

The effective Lagrangian relevant at low energies is the one of QCD
$+$ QED. The strength of the interaction is characterized by the two
coupling constants $g$ and $e$. In contrast to the Standard Model,
the SU(3)$\times$U(1) Lagrangian does contain mass terms: the quark and lepton
mass matrices $m_q$, $m_\ell$. The field basis may be chosen such that $m_q$
and $m_\ell$ are diagonal and positive. 

The two gauge fields behave in a 
qualitatively different manner: while the photons do not carry electric
charge, the gluons do carry colour. This difference is responsible for the fact
that the strong interaction becomes strong at low energies, while the
electromagnetic interaction becomes weak there, in fact remarkably weak:
the photons and leptons essentially decouple from the quarks and gluons.
For the QCD part, on the
other hand, perturbation theory is useful only at high energies. In the low
energy domain, the strong interaction is so strong that it 
confines the quarks and gluons. For the same reason, 
a term in the Lagrangian of the form $\sim \theta \,
G_{\mu\nu}\,\tilde{G}^{\mu\nu}$ (where $G_{\mu\nu}$ is the gluon field
strength) cannot a priori be dismissed, despite the fact
that it represents a total derivative: 
it generates an electric dipole moment in the neutron, for instance.
Conversely, the experimental fact that the dipole moment is smaller than 
$10^{-25}\,\mbox{ecm}$ implies that (in the basis where the quark mass matrix
is diagonal, real 
and positive) the vacuum angle $\theta$ must be tiny, so that the Lagrangian
is invariant under the discrete symmetries $P$, $C$ and $T$, to a very high
degree of accuracy. 

The resulting effective low energy theory is mathematically even more 
satisfactory than the 
Standard Model as such -- it does not involve scalar degrees of freedom and
has fewer free parameters. Remarkably, this simple theory 
must describe the
structure of cold matter to a very high degree of precision, once 
the parameters in the Lagrangian are known. It in particular 
explains the size of the atoms in terms of the scale
\bea a_{\hspace{-0.07em}\ind B}=\frac{4\,\pi}{e^2\,m_e}\co\nonumber\eea 
which only contains the two parameters $e$ and $m_e$ -- these are indeed known
to an incredible precision. Unfortunately, our ability to solve the QCD part
of the theory is rather limited -- in particular, we are
still far from being able to demonstrate on the basis of the QCD Lagrangian
that the strong interaction actually
confines colour. Likewise, our knowledge of the magnitude of the light quark
masses leaves to be desired -- we need to know these more accurately 
in order to test ideas that might lead to an understanding of the
mass pattern, such as the relations with the lepton masses that emerge from 
attempts at unifying the electroweak and strong forces.

\section{2.\hspace{0.6em}Symmetries of massless QCD}
In the following, I focus on the QCD part and switch the electromagnetic
interaction off.  It so happens that the interactions of $u,d,s$ with the
Higgs fields are weak, so that the masses $m_u,m_d,m_s$ are small. 
Let me first set these parameters equal to zero and, moreover, 
send the masses of the heavy quarks,
$m_c,m_b,m_t$ to infinity. In this limit, the theory becomes a theoreticians
paradise: the Lagrangian contains a single parameter, $g$. In fact, since 
the value of $g$ depends on the running scale used, the theory does not
contain any dimensionless parameter that would need to be adjusted to
observation. In principle, this theory fully specifies all dimensionless 
observables as pure numbers, while dimensionful quantities like masses or
cross sections can unambiguously
be predicted in terms of the scale $\Lambda_{\QCD}$ or the mass of the proton.
The resulting theory -- QCD with three massless flavours -- is among the
most beautiful quantum field theories we have. I find it breathtaking that, 
at low energies, nature reduces to this beauty, as soon as the dressing with
the electromagnetic interaction is removed and the Higgs condensate is 
replaced by one that does not hinder the light quarks, but is impenetrable for
$W$ and $Z$ waves as well as for heavy quarks.

The Lagrangian of the massless theory, ${\cal L}_{\QCD}^{\ind 0}$,
has a high degree of symmetry that originates in the fact that the
interaction among the quarks and gluons is flavour-independent and conserves
helicity: ${\cal L}_{\QCD}^{\ind 0}$ is invariant under independent flavour 
rotations
of the three right- and left-handed quark fields. These form the group
$G=\mbox{SU(3)}_\indR\times\mbox{SU(3)}_\indL$. The corresponding 16 currents
$V^\mu_i\qbar\,\gamma^{\,\mu}\frac{1}{2}\,\lambda_i q$ and 
$A^\mu_i=\qbar\,\gamma^{\,\mu}\gamma_5\frac{1}{2}\,\lambda_i q$ are conserved, so
that their charges commute with the Hamiltonian:
\bea [\,Q_i^{\indV},H_\QCD^{\ind 0}\,]=
[\,Q_i^{\indA},H_\QCD^{\ind 0}\,]=0\co
\hspace{2em}i=1,\,\ldots\,,8\fs\nonumber\eea
Vafa and Witten~\cite{Vafa Witten} have shown that the state of lowest energy
is necessarily invariant under the vector charges: 
$Q_i^{\indV}\rvac=0$. For the axial charges, however, there are the
two possibilities characterized in table 1.
\begin{table}[thb]
\begin{tabular}{|c|c|}
\hline
\rule[-0.7em]{0em}{1.9em}\rule{5.5em}{0em}$Q_i^{\indA}\rvac=0$
\rule{5.5em}{0em}&
\rule{5.5em}{0em}
$Q_i^{\indA}\rvac\neq0$\rule{5.5em}{0em}\\ \hline\rule{0em}{1.2em}
Wigner-Weyl realization of $G$&Nambu-Goldstone realization of $G$ \\
ground state is symmetric & ground state is asymmetric\\
\rule[-1em]{0em}{2em}
$\lvac\qbar_\indR q_\indL\rvac = 0$&$\lvac\qbar_\indR q_\indL\rvac \neq 0$
\vspace*{-0.5em}\\
ordinary symmetry & spontaneously broken symmetry\\
spectrum contains parity partners & spectrum contains Goldstone bosons\\
\rule[-0.7em]{0em}{0em}degenerate multiplets of $G$& 
degenerate multiplets of $\mbox{SU(3)}_{\indV}\subset G$\\
\hline\end{tabular}
\caption{Alternative realizations of the symmetry group
  $G=\mbox{SU(3)}_\indR\times\mbox{SU(3)}_\indL$.} 
\end{table}

The observed spectrum does not contain parity doublets. In the case
of the lightest meson, the pion, for instance, the lowest state with 
the same spin and flavour quantum numbers, but opposite parity is the 
$a_0(980)$. So, expe\-ri\-ment rules out the first possibility: 
for dynamical reasons that yet remain to be understood,
the state of lowest energy is an asym\-metric state. 
Since the axial charges
commute with the Hamiltonian, there must be eigenstates with the same energy
as the ground state:
\bea H^{\ind 0}_\QCD\, Q_i^{\indA}\rvac= Q_i^{\indA}\,H^{\ind 0}_\QCD\rvac
=0\fs\nonumber\eea
The spectrum must contain 8 states $Q_1^{\indA}\rvac,\ldots\,,
Q_8^{\indA}\rvac$  
with $E=\vec{P}=0$, describing massless particles, the Goldstone bosons of the
spontaneously broken symmetry. Moreover, these must carry spin 0, negative
parity and form an octet of SU(3). 
 
\section{3.\hspace{0.6em}Quark masses as symmetry breaking parameters}

Indeed, the 8 lightest hadrons,
$\pi^+,\pi^0,\pi^-,K^+,K^0,\bar{K}^0,K^-,\eta$,
do have these quantum numbers, but massless they
are not.
This has to do with the deplorable fact that we are not living in paradise: the
masses $m_u,m_d,m_s$ are different from zero and thus allow the left-handed
quarks to communicate with the right-handed ones. 
The full Hamilitonian is of the form
\bea  H_\QCD= H^{\ind 0}_\QCD+ H^{\ind 1}_\QCD\co\hspace{2em} 
H^{\ind 1}_\QCD=\int\!\!d^3\!x\;
\qbar_\indR m\, q_\indL+\qbar_\indL m^\dagger q_\indR\co\hspace{2em}
m=\left(\!\!\!\mbox{\begin{tabular}{ccc}\vspace*{-0.5em}$m_u\!\!\!$&
&\\\vspace*{-0.5em}&$\!\!\!m_d\!\!\!
$&\\&&$\!\!\!m_s$
\end{tabular}}\!\!\!\!\right)\fs\nonumber\eea
The quark masses may be viewed as symmetry breaking parameters: the 
QCD-Hamiltonian is only approximately symmetric under independent rotations of
the right- and left-handed quark fields, to the extent that these parameters
are small. Chiral symmetry is thus broken in two ways: 
\begin{itemize}\item spontaneously\hspace{3em} 
$\lvac\qbar_\indR q_\indL\rvac \neq 0$
\item explicitly\hspace{5.2em} $m_u,m_d,m_s\neq 0$\end{itemize}
The consequences of the fact that the explicit symmetry breaking is small may
be worked out by means of an effective field theory, ``chiral perturbation
theory'' \cite{Weinberg Physica,GL 1984,Jefferson}. In this context, 
the heavy quarks do not play
an important role -- 
as the corresponding fields are singlets under 
SU(3)$_{\indR}\times$SU(3)$_{\indL}$, we may include their contributions 
in the symmetric part of the Hamiltonian, 
irrespective of the size of their mass. 

Since the masses of the two lightest quarks are particularly small, the
Hamiltonian of QCD is almost exactly invariant under the subgroup 
SU(2)$_{\indR}\times$SU(2)$_{\indL}$. The ground state
spontaneously breaks that symmetry to the subgroup SU(2)$_{\indV}$ --
the good old isospin symmetry discovered in the thirties of the last
century \cite{Heisenberg}. The pions represent the
corresponding Goldstone bosons \cite{Nambu}, while the kaons and 
the $\eta$ remain
massive if the limit $m_u,m_d\rightarrow 0$ is taken at fixed $m_s$. 
In the following, I consider this framework and, moreover, ignore isospin
breaking, setting $m_u=m_d=\hat{m}$.

If SU(2)$_{\indR}\times$SU(2)$_{\indL}$ was an exact symmetry, the pions 
would be strictly massless. According to
Gell-Mann, Oakes and Renner \cite{GMOR}, the square of the pion mass is
proportional to the product of the quark masses and the quark condensate:
\bea\label{eq:GMOR} 
\al\al M^2_\pi\simeq\frac{1}{F_\pi^2}\times(m_u+m_d)\times |\lvac\, \ubar 
u \rvac|\fs\eea
The factor of proportionality is given by the pion decay constant
$F_\pi$. The term $m_u+m_d$ measures the explicit
breaking of chiral symmetry,  
while the quark condensate, 
\bdm \lvac\,\ubar u\rvac =\lvac\, \ubar_{\indR} 
u_{\ind L}\rvac +\mbox{c.c.}=
\lvac\,\dbar d\rvac\co\edm
is a measure of the spontaneous symmetry breaking: it may be viewed as an
order parameter and plays a role analogous to the spontaneous 
magnetization of a magnet.  

\section{4.\hspace{0.6em}Role of the quark condensate}

The approximate
validity of the relation (\ref{eq:GMOR}) was put to question by Stern and
collaborators \cite{KMSF}, who pointed out that there is no experimental
evidence for the quark condensate to be different from zero. Indeed,
the dynamics of the ground state of QCD is not understood -- it could resemble
the one of an antiferromagnet, where, for dynamical reasons, the
most natural candidate for an order parameter, the magnetization, 
happens to vanish. There are a number of theoretical reasons indicating 
that this scenario is unlikely:

(i) The fact that the pseudoscalar meson octet satisfies the Gell-Mann-Okubo 
formula remarkably well would then be accidental.

(ii) The value obtained for the quark condensate on the basis of QCD sum
rules, in particular for the baryonic correlation functions \cite{Ioffe}, 
confirms the standard picture.

(iii) The lattice values \cite{Lubicz} for the ratio $m_s/\hat{m}$
agree very well with the result of 
the standard chiral perturbation theory analysis
\cite{Leutwyler 1996}, corroborating this picture further.

Quite irrespective, however, of whether or not the scenario advocated 
by Stern et al.~is theoretically appealing, the issue can be subject to 
experimental test. In fact, significant progress has recently been achieved 
in this direction \cite{CGLPRL,Pislak}. The remainder of the talk
concerns this matter.

The Gell-Mann-Oakes-Renner formula is not exact. The expansion of $M_\pi^2$ 
in powers of $m_u,m_d$ contains an infinite  
sequence of contributions. The expansion starts with a term linear in the
quark masses: 
\bea\label{eq:Mexp} M_\pi^2=M^2-\frac{\lbar_3}{32\pi^2 F^2}\; M^4
+O(M^6)\co\hspace{2em}
M^2\equiv(m_u+m_d)\,B\fs \eea
The coefficient $B$ of the linear term is given 
by the value of 
$|\lvac\, \ubar u\rvac|/F_\pi^2$ in the limit $m_u,m_d\rightarrow 0$, and
$F$ is the value of $F_\pi$ in that limit. The Gell-Mann-Oakes-Renner formula
is obtained by dropping the higher order contributions. These are dominated by
the term of order $M^4$, which involves 
one of the coupling constants occurring in the
effective Lagrangian at order $p^4$. More precisely, the formula involves the
value of the running coupling constant $\ell_3$ at scale $\mu=M$, which 
logarithmically
depends on $M$. Expressed in terms of the corresponding intrinsic scale 
$\Lambda_3$, we have
\bea \lbar_3=\ln\frac{\Lambda_3^2}{M^2}\fs\eea
The symmetry does not determine the
numerical value of this scale. The crude estimates 
underlying the standard version of chiral perturbation theory \cite{GL 1984}
yield numbers in the range 
\bea \label{eq:Lambda3}0.2\;\mbox{GeV}<\Lambda_3<2\;\mbox{GeV}\fs\eea
The term of order $M^4$ is then very small compared to the one of order $M^2$,
so that the Gell-Mann-Oakes-Renner formula is obeyed very well.
Stern and collaborators investigate the more general framework, referred to as
``generalized chiral perturbation theory'', where
arbitrarily large values of $\lbar_3$ are considered. The
quartic term in eq.~(\ref{eq:Mexp}) can then take values comparable to the
``leading'', quadratic one. If so, the dependence of $M_\pi^2$ on the quark 
masses would fail to be approximately linear, even for values of $m_u$ and 
$m_d$ that are small compared to the intrinsic scale of QCD. A different 
bookkeeping for the terms 
occurring in the chiral perturbation series is then needed \cite{KMSF} 
-- the standard chiral power counting is adequate only if $\lbar_3$ is 
not too large.

\section{5.\hspace{0.6em}Quark mass dependence of $\mathbf{M_\pi}$ and 
$\mathbf{F_\pi}$} 
\begin{figure}[t] 
\psfrag{y}{}
\psfrag{Fpi}{\hspace{-3em}\raisebox{-0.4em}{$F_\pi/F$}}
\psfrag{Mpi}{\hspace{1em}\raisebox{0.6em}{$
M_\pi^2/M^2$}}
\psfrag{m}{\raisebox{-1.3em}{\hspace{-10em}$\hat{m}/m_s$}}

\mbox{\epsfysize=4.5cm \epsfbox{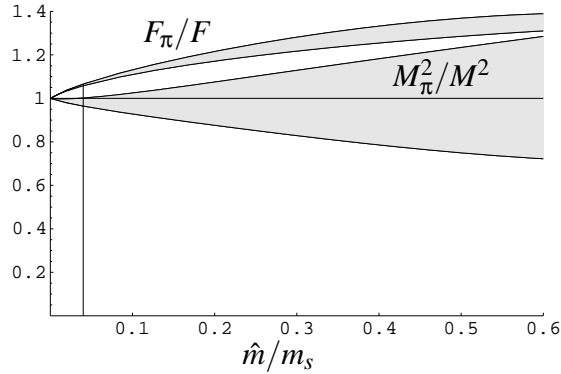} }\rule[-1em]{0em}{0em}

\caption{\label{fig:FM}Dependence of the ratios $F_\pi/F$ and $M_\pi^2/M^2$ on 
$\hat{m}=\frac{1}{2}(m_u+m_d)$. 
The strange quark mass is held fixed at the
physical value. The vertical line corresponds to the physical value of
$\hat{m}$.} 

\end{figure}

The behaviour of the ratio $M_\pi^2/M^2$ as a function of $\hat{m}$ is
indicated in fig.~\ref{fig:FM}, taken from ref.~\cite{bangalore}. 
The fact that the information about the
value of $\Lambda_3$ is very meagre shows up through very large uncertainties.
In particular, 
with $\Lambda_3\simeq 0.5 \,\mbox{GeV}$, the ratio $M_\pi^2/M^2$ would remain
close to 1, on the entire interval shown. Note that outside the range
(\ref{eq:Lambda3}),
the dependence of $M_\pi^2$ on the 
quark masses would necessarily exhibit strong curvature. 

The figure illustrates the fact 
that brute force is not the only way the very small values of
$m_u$ and $m_d$ observed in nature can be reached through numerical
simulations on a  
lattice. It suffices to equip the strange quark with the physical value of
$m_s$ and to measure the dependence
of the pion mass on $m_u,m_d$ in the region where $M_\pi$ is comparable to
$M_K$. A fit to the data based on eq.(\ref{eq:Mexp})  
should provide an extra\-po\-lation to the 
physical quark masses that is under good control\footnote{The logarithmic 
singularities occurring at
next-to-next-to-leading order are also known
\cite{Colangelo 1995} -- for a
detailed discussion, I refer to \cite{opus2}.}. Moreover, the fit would
allow a determination of the scale $\Lambda_3$ on the lattice. This is of 
considerable interest, because that scale
also shows up in other contexts, in the $\pi\pi$ scattering lengths, 
for example. For recent work in this direction, I refer
to \cite{Heitger,Durr}. 

For the pion decay constant, the expansion analogous to eq.~(\ref{eq:Mexp}) 
reads
\bea \label{eq:Fexp}F_\pi = F\left\{1+\frac{\lbar_4\,M^2}{16\pi^2 F^2}
+O(M^4)\right\}\co\hspace{2em}\lbar_4=\ln\frac{\Lambda_4^2}{M^2}\fs\eea
In this case, the relevant effective coupling constant
is known rather 
well: chiral symmetry implies that it also 
determines the
slope of the scalar form factor of the pion,
\bea F_s(t)=
\langle\pi(p')|\;\ubar u + \dbar d\,|\pi(p)\rangle=F_s(0)\left\{1+
\mbox{$\frac{1}{6}$}\,\langle r^2\rangle_s\,t+O(t^2)\right\}\fs\nonumber\eea
As shown in ref.~\cite{GL 1984}, the expansion of $\langle r^2\rangle_s$ in 
powers of $m_u,m_d$ starts with
\bea\label{eq:rs}\langle r^2\rangle_s=\frac{6}{(4\pi F)^2}
\left\{\lbar_4-\frac{13}{12}+O(M^2)\right\}\fs\eea
Analyticity relates the scalar form factor to the $I=0$ $S$--wave phase shift
of $\pi\pi$ scattering \cite{DGL}. Evaluating the relevant dispersion relation
with the remarkably accurate information about the phase shift that
follows from the Roy equations \cite{opus2}, one finds
$\langle r^2\rangle_s=0.61\pm 0.04\,\mbox{fm}^2$. Expressed in terms of the
scale $\Lambda_4$, this amounts to 
\bea\label{eq:Lambda4} \Lambda_4=1.26\pm 0.14\;\mbox{GeV}\fs\eea 
Fig.~\ref{fig:FM} shows that this information determines the quark mass
dependence of the decay constant to within rather narrow limits. 
The change in $F_\pi$ occurring 
if $\hat{m}$ is increased from the physical value to $\frac{1}{2}\,m_s$ is of 
the expected size, comparable to the difference between $F_K$ and $F_\pi$.
The curvature makes it evident
that a linear extrapolation from values of order $\hat{m}\sim 
\frac{1}{2}\,m_s$ down to the physical region is meaningless. 

\section{6.\hspace{0.6em}$\mathbf{\pi\pi}$ scattering}
The experimental test of the hypothesis that the quark
condensate represents the leading order parameter relies on the
fact that $\lvac\, \qbar q\rvac$ not only manifests itself in the dependence of
the pion mass on $m_u$ and $m_d$, but also in the low energy properties of the 
$\pi\pi$ scattering amplitude.

At low energies, the scattering amplitude is dominated by the contributions
from the $S$-- and $P$--waves, because the angular momentum barrier suppresses
the higher partial waves. Bose statistics implies that configurations with
two pions and $\ell=0$ are symmetric in flavour space and thus 
carry either isospin $I=0$ or $I=2$, so that there are
two distinct $S$--waves. For $\ell=1$, on the other hand, the configuration
must be antisymmetric in flavour space, so that there is a single $P$--wave,
$I=1$. If the relative momentum tends to zero, only the $S$--waves contribute,
through the corresponding scattering lengths $a_0^0$ and $a_0^2$ (the
lower index refers to angular momentum, the upper one to isospin).

As shown by Roy \cite{Roy}, analyticity, unitarity and
crossing symmetry subject the partial waves to a set of coupled integral
equations. These equations involve two subtraction constants, which may be
identified with the two $S$--wave scattering lengths $a_0^0$, $a_0^2$. 
If these two constants are given, the Roy equations allow us to calculate the
scattering 
amplitude in terms of the imaginary parts above 800 MeV and the available
experimental information suffices to evaluate the relevant dispersion
integrals, to within small uncertainties \cite{ACGL}. In this sense, 
$a_0^0$, $a_0^2$ represent the essential parameters in low energy
$\pi\pi$ scattering. 

\begin{figure}[t]
\psfrag{a0}{$a_0^0$}
\psfrag{a2}{$a_0^2$}
\includegraphics[width=6cm,angle=-90]{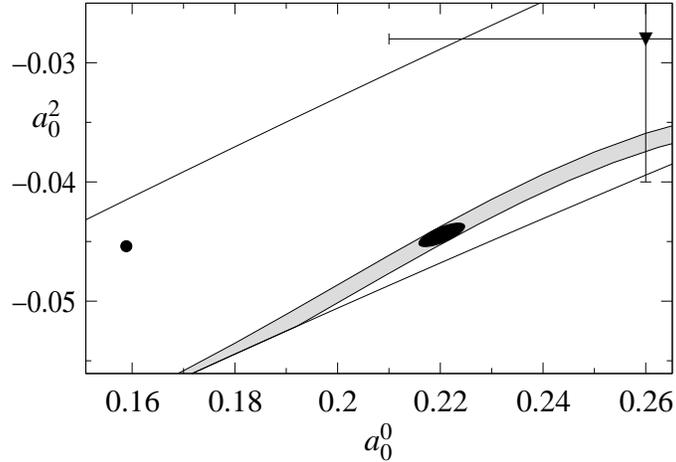}
\caption{\label{fig:aellipse} $S$--wave scattering
  lengths. The Roy
  equations only admit solutions in the ``universal band'',  
  spanned by the two tilted lines. The dot indicates Weinberg's leading order
  result, while the small ellipse includes the higher order corrections,
  evaluated in the framework of standard chiral perturbation theory. 
  In the generalized scenario, there is no prediction for $a_0^0$, but there
  is a correlation 
  between $a_0^0$ and $a_0^2$, shown as a narrow strip. The 
  triangle with error bars indicates the phenomenological
  range permitted by the old data, $a_0^0=0.26\pm 0.05$,
  $a_0^2=-0.028\pm0.012$ \protect\cite{Froggatt:1977hu}. }
\end{figure}

As a general consequence of the hidden symmetry, Goldstone bosons
of zero momentum cannot interact with one another. Hence the
scattering lengths $a_0^0$ and $a_0^2$ must vanish in the symmetry limit,
$m_u,m_d\rightarrow 0$. These quantities thus also measure the explicit
symmetry breaking generated by the quark masses, like $M_\pi^2$. In fact, 
Weinberg's low energy theorem~\cite{Weinberg 1966} states that, 
to leading order of the expansion in powers of $m_u$ and $m_d$,
the scattering lengths are proportional to $M_\pi^2$, the factor of
proportionality being fixed by the pion decay constant:\footnote{The
standard definition of the scattering length 
corresponds to $a_0/M_\pi$. It is not suitable in the present context, 
because it differs from the invariant 
scattering amplitude at threshold by a kinematic factor that diverges in the
chiral limit.}
\bea\label{eq:Weinberg} a_{0}^0=\frac{7 M_\pi^2}{32 \,\pi \, F_\pi^2}+
O(\hat{m}^2)\co
\hspace{1.3em}
a_{0}^2=-\frac{M_\pi^2}{16 \,\pi \,
  F_\pi^2}+O(\hat{m}^2)\fs\eea
Chiral symmetry thus provides the missing element: 
in view of the Roy equations,
Weinberg's low energy theorem fully determines the low energy behaviour of the
$\pi\pi$ scattering amplitude. The prediction (\ref{eq:Weinberg}) corresponds 
to the dot on the left of fig.~\ref{fig:aellipse}.

The prediction is of limited accuracy, 
because it only holds to leading order of the expansion in powers of the quark
masses. In the meantime, the chiral perturbation series of the scattering
amplitude has been worked out to two loops \cite{BCEGS}.
At first nonleading order of the expansion in powers
of momenta and quark masses,  the scattering amplitude
can be expressed in terms of $F_\pi$, $M_\pi$ and the 
coupling constants $\ell_1,\ldots\,,\ell_4$ that occur in the derivative
expansion of the effective Lagrangian at order $p^4$. The terms $\ell_1$ and
$\ell_2$ manifest themselves in the energy dependence 
of the scattering amplitude and can thus be determined phenomenologically. 
As discussed in section 5, the coupling constant $\ell_4$ is known rather
accurately from the dispersive analysis of the scalar form factor.
The crucial term is $\ell_3$ -- the range considered for this
coupling constant makes the difference between standard and generalized chiral
perturbation theory. In the standard framework, where the relevant scale
is in the range (\ref{eq:Lambda3}), one finds that the 
leading order result is shifted into the small ellipse shown
in fig.~\ref{fig:aellipse}, which corresponds to \cite{ABT,CGL}: 
\bea\label{eq:a0a2} 
a_0^0=0.220\pm0.005\co\hspace{2em}a_0^2=-0.0444\pm0.0010\fs\eea 
The numerical value quoted includes the higher order corrections 
(in the standard framework, the contributions
from the corresponding coupling constants are tiny).

The corrections from the higher order terms in the Gell-Mann-Oakes-Renner
relation can only be large if the estimate (\ref{eq:Lambda3}) for 
$\Lambda_3$ is totally wrong. As pointed out long ago \cite{GL 1983}, 
there is a low energy theorem
that holds to first nonleading order and relates the $S$--wave scattering
lengths to the scalar radius: 
\bea\label{eq:one loop}2a_0^0-5a_0^2=
\frac{3\,M_\pi^2}{4\pi F_\pi^2}\left\{1+
\frac{1}{3}\,M_\pi^2 \rs+\frac{41\,M_\pi^2}{192 \,\pi^2 F_\pi^2}\right\}
 + 
O(\hat{m}^3)\fs\eea
In this particular combination of scattering lengths, the term $\ell_3$ drops
out.  The theorem thus correlates the two scattering lengths, independently
of the numerical value of $\Lambda_3$. The correlation holds 
both in standard and generalized chiral perturbation theory. The corrections
occurring in eq.~(\ref{eq:one loop}) 
at order $\hat{m}^3$  have also been worked out. These are responsible for the
fact that the narrow strip, which represents the correlation in 
fig.~\ref{fig:aellipse}, is slightly 
curved.

\section{7.\hspace{0.6em}Impact of the new $\mathbf{K}$ decay data}
The final state interaction theorem implies that the phases of the form 
factors relevant for the decay $K\rightarrow \pi\pi e\nu$ are determined by
those of the $I=0$ $S$--wave and of the $P$--wave of elastic 
$\pi\pi$ scattering, respectively.
Conversely, the analysis of the final state distribution observed in this 
decay yields a measurement of the phase difference
$\delta(s)\equiv\delta_0^0(s)-\delta_1^1(s)$, in the region 
$4M_\pi^2<s<M_K^2$. As discussed above, the Roy equations determine
the behaviour of the phase shifts in terms of the two
$S$--wave scattering lengths. Moreover, in view of the correlation between
the two scattering lengths, $a_0^2$ is determined by $a_0^0$, so that the
phase difference $\delta(s)$ can be calculated as a function of $a_0^0$ 
and $q$, where $q$ is the c.m.~momentum
in units of $M_\pi$, $s= 4M_\pi^2(1+q^2)$.
In the region of
interest ($q<1$, $0.18<a_0^0<0.26$), the prediction reads 
\bea\label{eq:delta(s)} &&\delta_0^0-\delta_1^1=
\frac{q}{\sqrt{1+q^2}}\,( a_0^0+q^2\,b+q^4\,c+q^6\,d)\pm e\\
&&b= 0.2527+0.151\,\Delta a_0^0+1.14\,(\Delta a_0^0)^2 +
35.5\,(\Delta a_0^0)^3\co\no
&&c=0.0063-0.145\,\Delta a_0^0\co\hspace{2em}
d=-0.0096\co\nonumber\eea
with $\Delta a_0^0=a_0^0-0.22$.
The uncertainty in this relation mainly stems from the experimental input used
in the Roy equations and is not sensitive to $a_0^0$:
\bea\label{eq:errordelta} e= 0.0035 \,q^3+0.0015\,q^5\fs\eea
\begin{figure}[thb]
\leavevmode

\centering
\includegraphics[width=10cm]{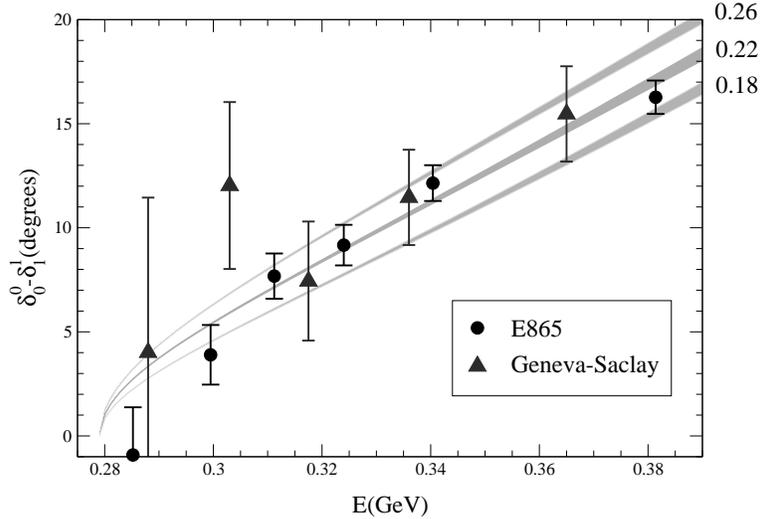}
\caption{\label{fig:deltaKl4} Phase relevant for the decay $K\rightarrow
  \pi\pi e\nu$. The three bands correspond to the 
three indicated values of the $S$--wave scattering length $a_0^0$. The
uncertainties are dominated by those from the experimental input used
in the Roy equations. The triangles are the data points of
ref.~\protect\cite{rosselet}, while the full circles represent the 
E865 results \protect\cite{Pislak}.}
\end{figure}

The prediction (\ref{eq:delta(s)}) is illustrated
in fig.~\ref{fig:deltaKl4}, where the energy dependence of  
the phase difference is shown for $a_0^0=0.18$, $0.22$ and
$0.26$. The width of the corresponding bands indicates the uncertainties,
which according to (\ref{eq:errordelta})
grow in proportion to $q^3$ -- in the range shown, they
amount to less than a third of a degree. 

The figure shows that the data of ref.~\cite{rosselet} barely 
distinguish between the three values of $a_0^0$ shown.  
The results of the E865 experiment at Brookhaven \cite{Pislak} are
significantly more precise, however. The best fit to these data is obtained
for $a_0^0=0.218$, with $\chi^2= 5.7$ for 5 degrees of freedom. This
beautifully confirms the value in eq.~(\ref{eq:a0a2}), 
obtained on the basis of standard chiral perturbation theory. 
There is a marginal
problem only with the bin of lowest energy: the corresponding scattering
lengths are outside the region
where the Roy equations admit solutions. In view of the
experimental uncertainties attached to that point, this discrepancy is without
significance: the difference between the central experimental value and 
the prediction amounts to 1.5 
standard deviations. Note also that the old data are perfectly consistent with
the new ones: the overall fit
yields $a_0^0=0.221$ with $\chi^2= 8.3$ for 10 degrees of freedom.

\begin{figure}[t]
\leavevmode
\centering
\includegraphics[width=8cm]{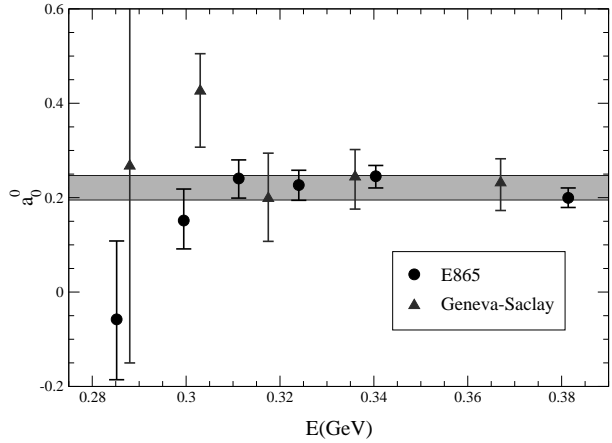}
\caption{\label{fig:aKe4}  $K_{e_4}$ data on the scattering length $a_0^0$.  
The triangles are the data points of
ref.~\protect\cite{rosselet}, while the full circles represent the 
E865 results \protect\cite{Pislak}. The horizontal band indicates the 
statistical average of the 11 values for $a_0^0$ shown in the figure.}
\end{figure}

The relation  (\ref{eq:delta(s)}) can be inverted, so
that each one of the values found for the phase difference yields
a measurement of the scattering length $a_0^0$.  
The result is shown in fig.~\ref{fig:aKe4}.
The experimental errors are remarkably small. It is not unproblematic,
however, to treat the data collected in the 
different bins as statistically independent: in the presence of correlations,
this procedure underestimates the actual uncertainties. Also, since the phase
difference rapidly rises with the energy, the binning procedure may introduce
further uncertainties. To account for this, the final result given in
ref.~\cite{CGLPRL},
\bea\label{eq:final result} a_0^0=0.221\pm 0.026\co\eea
corresponds to the 95\% confidence limit -- in effect, this amounts to
stretching the statistical error bar by a factor of two.

We may translate the result into an estimate for the magnitude 
of the coupling constant $\lbar_3$: the range (\ref{eq:final result})  
corresponds to $|\lbar_3|\,\lsim\, 16$.
Although this is a coarse estimate, it implies that the
Gell-Mann-Oakes-Renner relation does represent a decent approximation: 
 more than 94\% of the pion mass stems from the first term 
 in the quark mass expansion (\ref{eq:Mexp}), 
 i.e.~from the term that originates in the quark condensate.
This demonstrates that there is no need for a reordering of the chiral
perturbation series based on SU(2)$_{\indR}\times$SU(2)$_{\indL}$. In that
context, the generalized scenario has served its purpose and can now be
dismissed.    

A beautiful experiment is under way at CERN \cite{Nemenov}, 
which exploits the fact that $\pi^+\pi^-$ atoms decay into a pair of neutral
pions, through the strong transition
$\pi^+\pi^-\!\rightarrow\!\pi^0\pi^0$. Since 
the momentum transfer nearly vanishes, only the scattering lengths are
relevant: at leading order in isospin breaking, the transition amplitude is 
proportional to $a_0^0\!-\!a_0^2$. The corrections at
next--to--leading order are now also known~\cite{GLR}. Hence 
a measurement of the
lifetime of a $\pi^+\pi^-$ atom amounts to a measurement of
this combination of scattering lengths. At the planned accuracy of 10\% for
the lifetime, the experiment will yield a measurement of the scattering 
lengths to 5\%, thereby subjecting chiral perturbation theory to a very
sensitive test.

\end{document}